\documentclass{article}
\usepackage{amsmath,epsfig}
\usepackage[preprint]{spconfa4}
\usepackage{multirow} 
\usepackage[hang,flushmargin]{footmisc}

\copyrightnotice{ \\ \textbf{978-1-7281-1331-9/20/\$31.00~\copyright 2020 IEEE}}

\let\OLDthebibliography\thebibliography
\renewcommand\thebibliography[1]{
  \OLDthebibliography{#1}
  \setlength{\parskip}{0pt}
  \setlength{\itemsep}{0pt plus 0.3ex}
}

\begin{document}\sloppy

\def\x{{\mathbf x}}
\def\L{{\cal L}}

\title{Attention-based network for low-light image enhancement} 
%

\name{Cheng Zhang$^{1,\ast}$\thanks{\noindent $^{\ast}$ Equal contributions.
This work was partially supported by NSFC (61901384, 61871328), ARC(DP160100703) and National Engineering Laboratory for Integrated Aero-Space-Ground-Ocean Big Data Application Technology. Corresponding author: Yu Zhu.}, 
Qingsen Yan$^{2,\ast}$, Yu Zhu$^{1}$, Xianjun Li$^{1}$, Jinqiu Sun$^{1}$, Yanning Zhang$^{1}$}

\address{$^{1}$Northwestern Polytechnical University, $^{2}$The University of Adelaide \\ 
zhangcheng233@mail.nwpu.edu.cn, ynzhang@nwpu.edu.cn
}


\maketitle

%
\begin{abstract}
The captured images under low-light conditions often suffer insufficient brightness and notorious noise.
Hence, low-light image enhancement is a key challenging task in computer vision. A variety of methods have been proposed for this task, but these methods often failed in an extreme low-light environment and amplified the underlying noise in the input image.
To address such a difficult problem, this paper presents a novel attention-based neural network to generate high-quality enhanced low-light images from the raw sensor data. 
Specifically, we first employ attention strategy (\textit{i.e.} spatial attention and channel attention modules) to suppress undesired chromatic aberration and noise.
The \textit{spatial attention module} focuses on denoising by taking advantage of the non-local correlation in the image.
The \textit{channel attention module} guides the network to refine redundant colour features.
Furthermore, we propose a new pooling layer, called \textit{inverted shuffle layer}, which adaptively selects useful information from previous features.
Extensive experiments demonstrate the superiority of the proposed network in terms of suppressing the chromatic aberration and noise artifacts in enhancement, especially when the low-light image has severe noise.
\end{abstract}
\section{Introduction}
\label{sec:intro}

Image brightness is determined by irradiance of the scene and the camera setting.
Images captured in the insufficient irradiance environment usually suffer multiple degradations, such as poor visibility, low contrast, unexpected noise.
Unfortunately, these images inevitably exist in our daily life, especially at night or indoors.
Such images have poor visual effects and are difficult to use as input for other visual tasks such as target detection and recognition.
Although the auto-exposure mechanism (\textit{e.g.} ISO, shutter, flashlight, etc.) can correctly enhance image brightness, it often causes other unexpected artifacts (\textit{e.g.} noise, blurring, over-saturation, etc.).
Hence, restoring normally exposed high-quality images from low-light images plays an important role in practical application.

\begin{figure}[t]
  \centering
  \begin{minipage}[b]{0.49\linewidth}
    \includegraphics[width=1.0\textwidth]{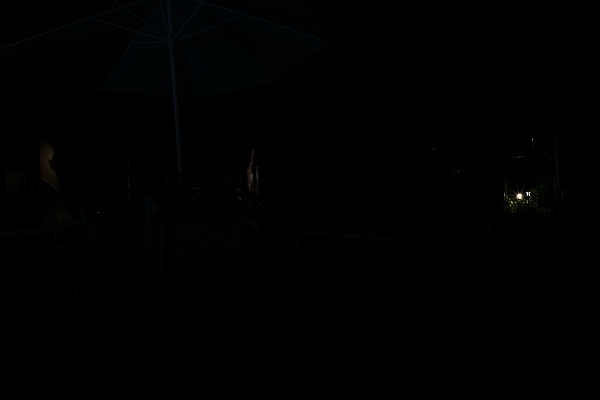}
    \centerline{(a) Low-light image}
  \end{minipage}
  \hfill
  \begin{minipage}[b]{0.49\linewidth}
    \includegraphics[width=1.0\textwidth]{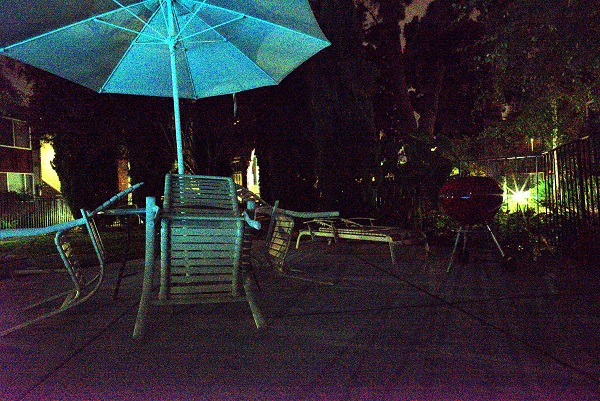}
    \centerline{(b) 300x scaling }
  \end{minipage}
  \hfill
  \begin{minipage}[b]{0.49\linewidth}
    \includegraphics[width=1.0\textwidth]{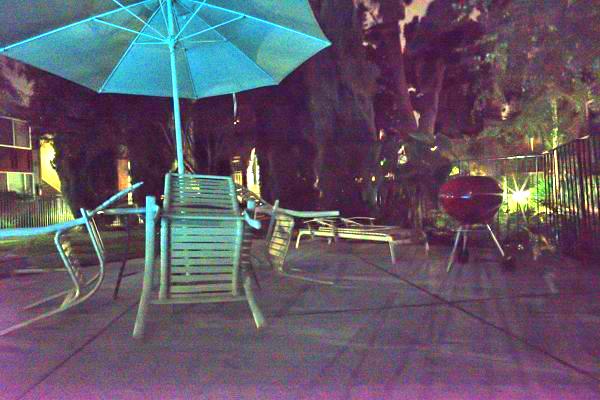}
    \centerline{(c) LIME}
  \end{minipage}
  \hfill
  \begin{minipage}[b]{0.49\linewidth}
    \includegraphics[width=1.0\textwidth]{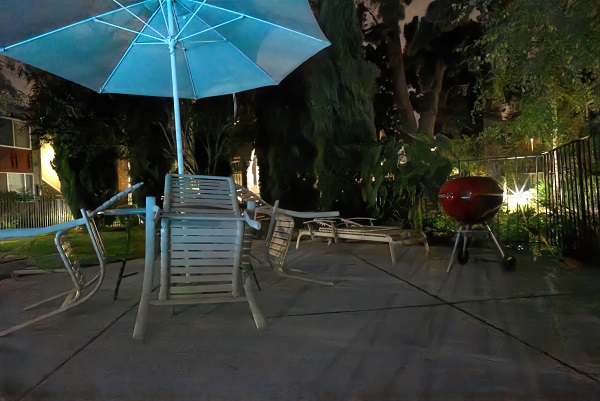}
    \centerline{(d) Ours}
  \end{minipage}
  \caption{(a) a raw image captured in extreme low-light environment; (b) 300x intensity scaling of (a); (c) the result of LIME\cite{guo2016lime}; (d) Ours enhanced image from (a). It is obvious that there are serious noise and color artifacts in exist methods.}
  \label{fig:introduce}
\end{figure}

Recent years, numbers of methods \cite{guo2019toward,shen2017msr,chen2018image} have been proposed for restoring low-light images, but there is still lots of room to be improved. Fig. \ref{fig:introduce} illustrates the limitations of existing methods. The reason why these methods like \cite{shen2017msr,chen2018image} failed in extreme low-light environment is that they focus on increasing the contrast and brightness, while ignoring the influence of serious noise, which results in noise amplification. Although the networks proposed by~ \cite{chen2018learning,maharjan2019improving} can generate high-quality images with processing noise and increasing brightness simultaneously, there remains color artifacts which affect visual equality seriously. 

To solve the problem of noise amplification and color artifacts in previous works, we propose an end-to-end network based on attention mechanism for processing low-light images.
We observed that a larger receptive field is the key to reduce color artifacts in low-light images since a wider range of information can guide the network to learn what it should be when suffering serious noise. 
Different from simply stacking residual layers to enlarge receptive field \cite{maharjan2019improving}, we design a new block, called mixed attention block, to effectively fuse local and global features in our network.
The proposed mixed attention block (\textit{i.e.} channel attention and spatial attention modules) effectively suppresses undesired chromatic aberration and noise.
The channel attention module guides the network to refine redundant color features. The spatial attention module focuses on denoising by taking advantage of the non-local correlation in the image.
In addition, considering that the max pooling layer often brings about information loss, we employ a new pooling strategy, called Inverted Shuffle Layer (ISL), to adaptively select important information from feature maps.
Overall, our contributions are in three folds: 
\begin{itemize}
  \item We propose an end-to-end network based on mixed attention block to obtain normally exposed high-quality and noise-free images. The mixed attention block includes spatial attention and channel attention, which can take into account local and global information.
  \item To reduce the information loss and select useful features flexibly, we employ the ISL to replace the max pooling layer.
  \item We evaluate our method on the SID dataset, and the experimental results demonstrate that our method achieves state-of-the-art performance.
\end{itemize}

\section{RELATED WORK}

Obtaining visually-friendly color images from raw images usually requires denoising, enhancement, etc. Therefore, we provide a literature review of the two tasks here.

\subsection{Image Denoising}
Image denoising is a fundamental task in computer vision.
In order to recover clear images from noisy ones, a variety of image priors have been proposed in the past years, including sparsity, low-rank, and self-similarity. Many based on image priors methods have made great progress in image denoising, such as BM3D \cite{dabov2007image}, WNNM \cite{gu2014weighted}. With the development of deep learning,
researchers have applied deep neural networks to image denoising in recent years. For example, DnCNN \cite{zhang2017beyond} trained a deep residual network and used batch normalization layers to speed up the training process. CBDNet \cite{guo2019toward} considered the noise in the whole process of imaging and adopted the U-net architecture with a sub-network to estimate noise levels for improving denoising performance.

\subsection{Low-light Image Enhancement}

Image enhancement has a long history in low-level vision. Histogram equalization and gamma correction are simple but classical methods that usually been applied to increase image contrast. It is obvious that those methods only adjust the contrast of the whole image globally, ignoring local brightness differences.

With the rapid development of deep learning
\cite{yan2020deep,yan2019attention,gong2018learning},
many methods are based on the Retinex theory that assumes an image can be decomposed into illumination and reflectance components. Shen \emph{et al.} \cite{shen2017msr} regarded multi-scale Retinex as a feedforward convolutional neural network and proposed MSR-net to learn a mapping between dark and bright images. RetinexNet \cite{wei2018deep} is another method inspired by the Retinex theory, which first decomposes the image into illuminance and reflection components with decompose subnetwork, and then performs image enhancement. Wang \emph{et al.} \cite{wang2019underexposed} proposed using the network to estimate the illuminance component of the image, and used the illuminance constraint and the prior in the loss function. Chen \emph{et al.} \cite{chen2018learning} considered the low-light image enhancement directly from the raw data. They created the SID dataset and obtained enhanced images in sRGB space with trained U-net. Paras \emph{et al.} \cite{maharjan2019improving} proposed to use residual learning to improve the enhancement performance for decreasing the amount of parameters and alleviating the impact of chromatic aberration.

\begin{figure*}[t]
\begin{minipage}[b]{1.0\linewidth}
 \centering
  \includegraphics[width=1\textwidth,trim=3 3 3 3,clip]{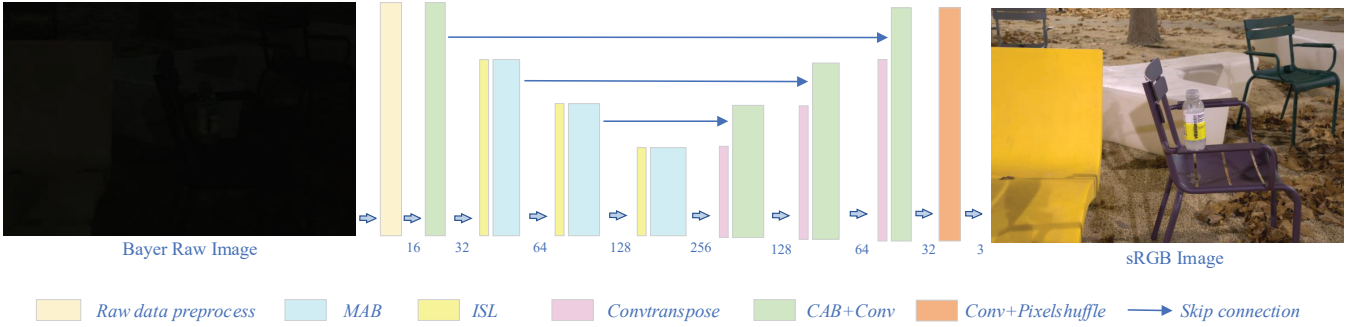}
\end{minipage}
\caption{Overview of our network structure. The network has a data pre-process layer and multiple proposed layers with the form of U-net.}
\label{fig:flow}
\end{figure*}

\section{Method}

Low-light image enhancement from the camera sensor is a complicated problem.
Traditional Image Signal Process (ISP) method consists of a series of subtasks (\textit{e.g.} white balance, demosaicking, denoising, etc.), however, it results in a high noise level and less vivid color \cite{liang2019cameranet}.
To mitigate these problems, we propose a novel Attention-based Low-light image Enhancement Network (ALEN) which directly converts raw image to color image. 

For a given low-light raw image $I_r$, the estimated color image $I_e$ can be defined as:
\begin{equation}
  I_e = F(I_r,\theta)
\end{equation}
where $F$ denotes the proposed network, and $\theta$ represents the parameters of the network. 
We present the details of the architecture and loss function in the following.

\subsection{Network Architecture}
As shown in Fig. \ref{fig:flow}, our network is in the form of U-net, which demonstrates its advantages in many tasks.
The proposed network consists of encoder, decoder and skip connections. 
In the raw data preprocess layer, inspired by multi-exposure, the image is multiplied by different amplification factors as input.
In the encoder part, we employ several mixed attention blocks and ISLs to obtain semantic features.
The mixed attention block which contains channel attention and spatial attention is beneficial to remove the color artifacts caused by multiplying amplification ratio.
On the other hand, the decoder part adopts multiple convolutional layers and transposed convolution to restore high-resolution features from semantic features.
Finally, the estimated image is obtained after a pixel shuffle operation from a 12-channel feature map.

\begin{figure}[t]
  \begin{minipage}[b]{0.88\linewidth}
    \centering
    \includegraphics[width=0.8\textwidth,trim=20 20 20 20]{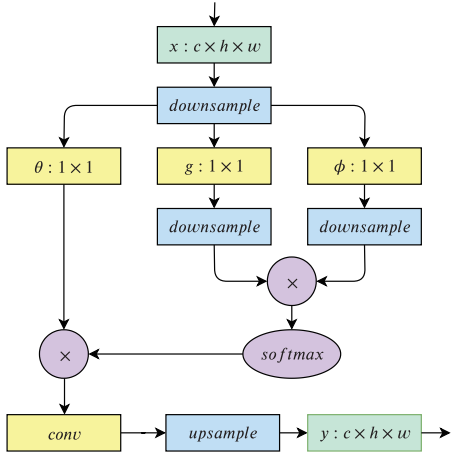}
    \centerline{(a) non-local operation}
  \end{minipage}
  \begin{minipage}[b]{0.45\linewidth}
    \centering
    \includegraphics[width=0.8\textwidth,trim=2 2 2 2]{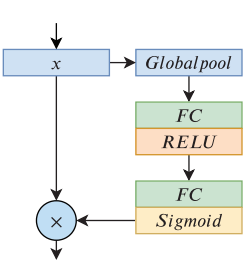}
    \centerline{(b) channel attention block}
  \end{minipage}
  \hfill
  \begin{minipage}[b]{0.46\linewidth}
    \centering
    \includegraphics[width=0.7\textwidth,trim=2 2 2 2]{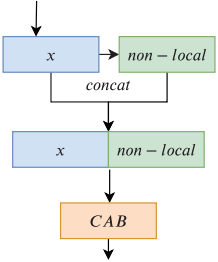}
    \centerline{(c) mixed attention block}
  \end{minipage}
  \caption{Structure of attention block}
  \label{fig:att}
\end{figure}

\textbf{Channel Attention Block}
Since the feature map of each channel has different contributions to the following network.
Therefore, we introduce a channel attention strategy to extract more useful information for low-light image enhancement. 
The structure of channel attention is illustrated in Fig. \ref{fig:att}(c). 
Like SEblock \cite{hu2018squeeze}, we first use a global average pooling layer to get a representative value in each channel.
Then we use two fully connection layers and activate functions to learn the significance between channels. 
The first fully connection layer is followed by a ReLU activate function and the second activate function is the Sigmoid function. 
The proposed channel attention block is well-motivated, it not only removes harmful features of inputs but also highlights the favorable color information.

\textbf{Non-local Operation}
The larger receptive field is critical in many computer vision tasks. But convolution operation is capable of processing a local neighborhood in space, thus capturing long-range information from feature map demand repeating local operation which is computationally inefficient.
Non-local operation is one way to tackle the above issue in recent years.
From \cite{wang2018non}, non-local operation can be expressed as 
\begin{equation}\label{equ:non-local}
  y_i=\sum_{j=1}^{N}\frac{h(x_i,x_j)}{C(x_i)}(G(x_i)),
\end{equation}
where $i$ is the query position, and $j$ is one of $N$ possible positions in feature map. $G(x_i)$ denotes the transform of $x_i$. $h(x_i,x_j)$ represents the relationship of $x_i$ and $x_j$. $C(x_i)$ is a normalization factor that is the sum of all $h(x_i,x_j)$ as
\begin{equation}
  C(x_i) = \sum_{j=1}^{N}h(x_i,x_j).
\end{equation}

Non-local operation aimed at strengthening the feature representation capability of the network.
Equation \ref{equ:non-local} shows that the result of non-local operation is a weighted sum of features at all positions. Thus, to utilize non-local operation makes the network have a global receptive field via aggregating different position information in a feature map. It is significant to correct the color and suppress noise, especially in a low-light environment, since a wider range of information is able to guide the network to learn what it should be in a seriously degraded scene. The structure is illustrated in Fig. \ref{fig:att}(a). In practical applications, non-local operation usually occupies a large memory and computation. Therefore, we adopt to downsample the feature to reduce computational complexity. 

\textbf{Mixed Attention Block}
As discussed above, channel attention block can model the interdependence between channels, while non-local operation can aggregate information from different positions in a feature map. 
In order to obtain better feature representation, we combine two attention blocks to a mixed attention block. Fig. \ref{fig:att}(b) illustrates the structure of mixed attention block. In this block, we first employ non-local operation to obtain features with a wider range of information in the spatial domain. Then we concatenate them and feed the concatenated features to channel attention block to generate final feature representation. With the mixed attention block, the network can make full use of information from different channels and positions in the feature map to produce a more flexible structure. 

\textbf{Inverted Shuffle Layer}
As we all know, the pooling layer usually appears in neural networks for reducing the computation with smaller feature sizes. However, pooling operation usually abandons useful information in the forward process whether it is max pooling or average pooling. Inspired by pixel shuffle in \cite{shi2016real}, we proposed a new pooling operation, named ISL, which includes inverted shuffle and convolution operation. After an inverted shuffle operation, the size of the feature map reduces to half of the original and the number of channels quadruples. Convolution layer with $1\times 1$ kernels is performed after the inverted shuffle, which plays a role in selecting useful information while compressing the number of channels. 
In general, ISL not only has the effect of reducing the computation as a pooling layer but also makes the network more flexible to select features.

\subsection{Loss Function}
In our network, we combine L1 loss and SSIM loss with a weight which usually appears in image restoration methods. The loss function of our method can be expressed as
\begin{equation}
  L = \sum_{i=1}^{N}\alpha L_{1}^{i}+(1-\alpha)L_{ssim}^{i}
\end{equation}
where $L_{1}^{i}$ is pixel wise L1 loss, and $L_{ssim}^{i}$ denotes SSIM loss, $\alpha$ is the weight to balance L1 loss and SSIM loss. Note that we set $\alpha = 0.85$ in the train process.

\begin{figure*}[h]s
  \begin{minipage}[b]{0.24\linewidth}
   \centering
   \includegraphics[width=1.0\textwidth]{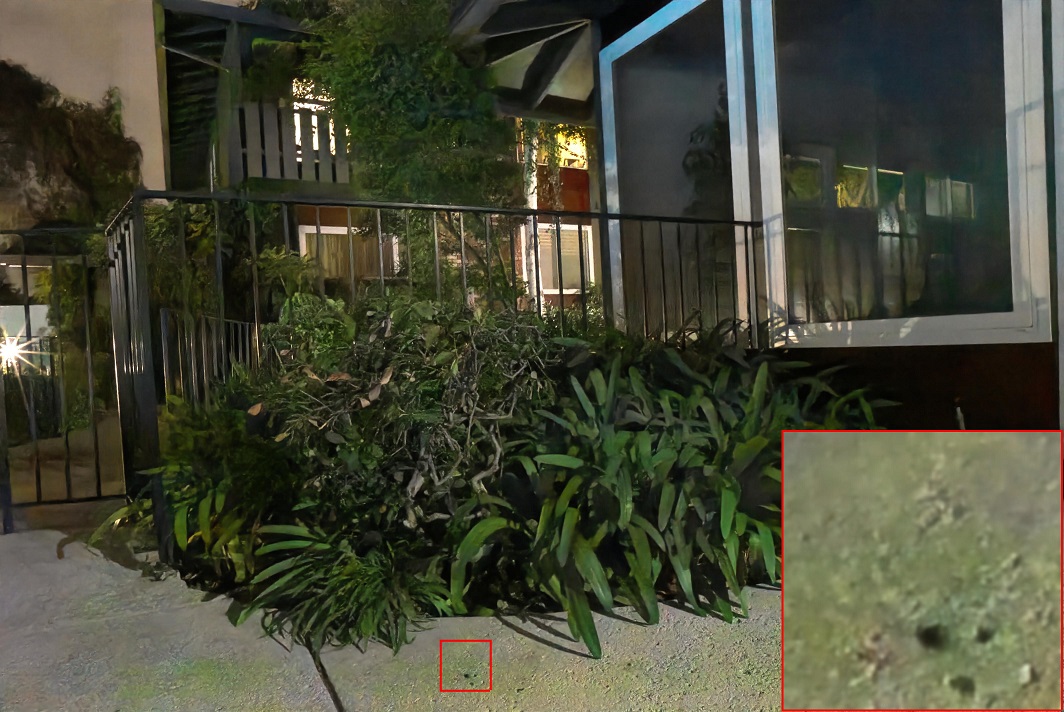}
    \includegraphics[width=1.0\textwidth]{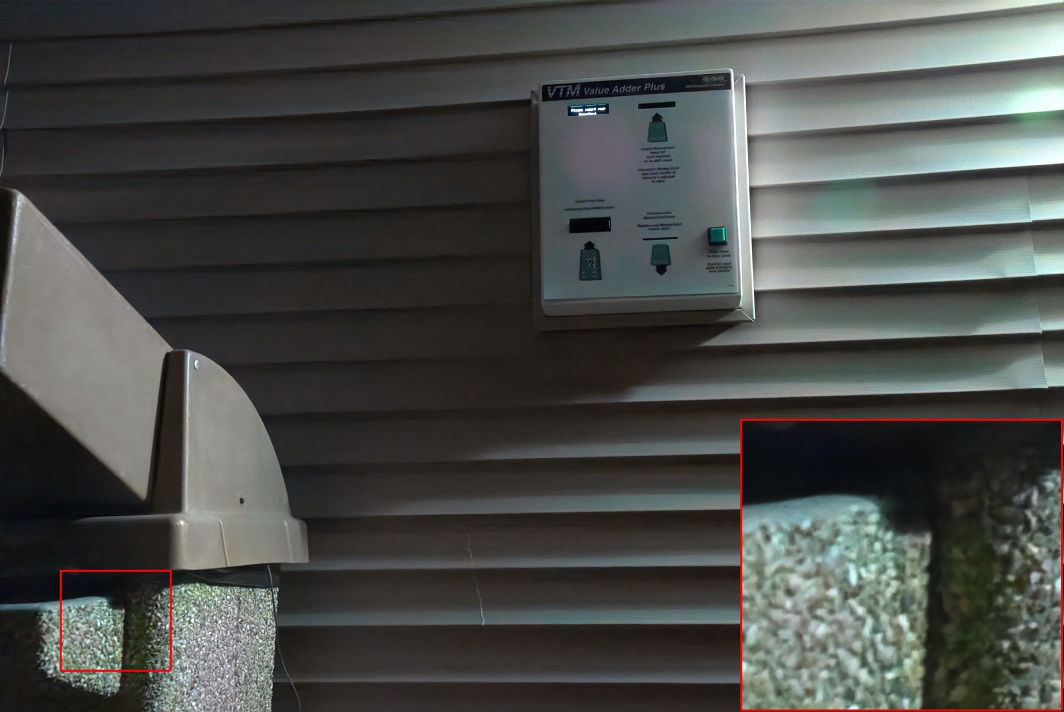}
    \includegraphics[width=1.0\textwidth]{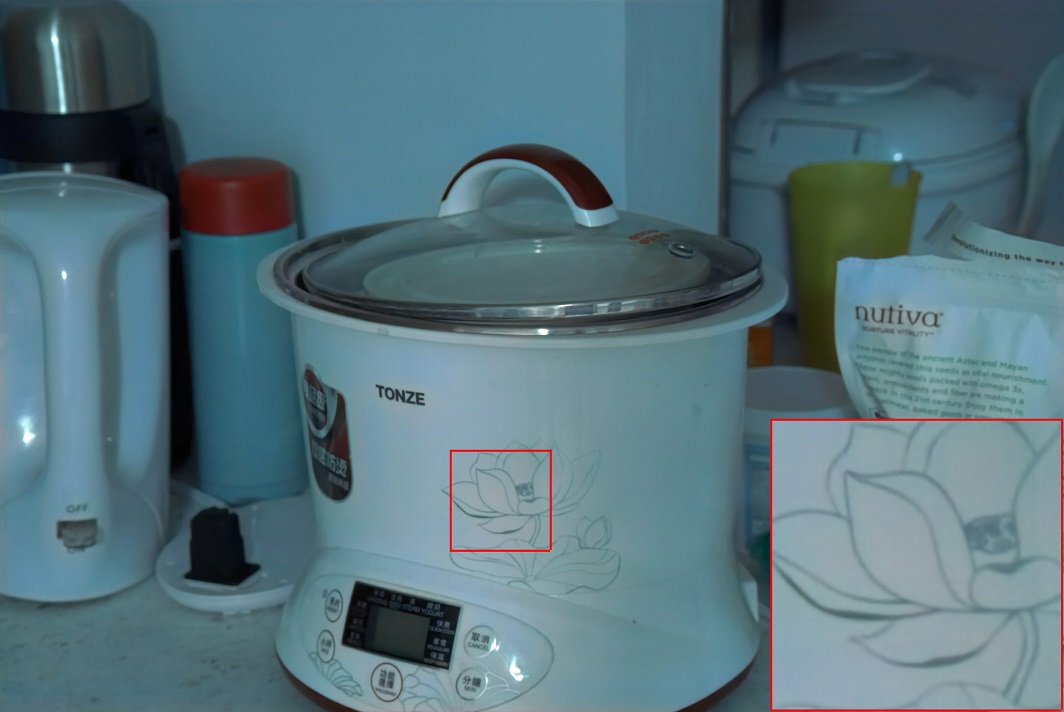}
    \centerline{(a) SID }
  \end{minipage}
  \hfill
  \begin{minipage}[b]{0.24\linewidth}
   \centering
   \includegraphics[width=1.0\textwidth]{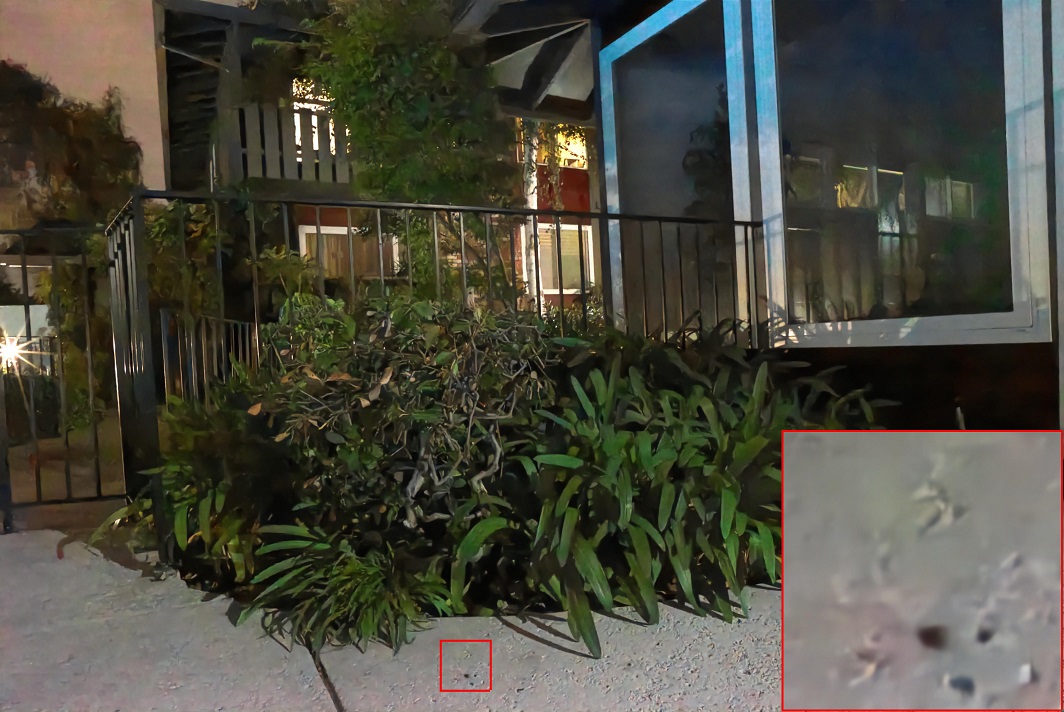}
   \includegraphics[width=1.0\textwidth]{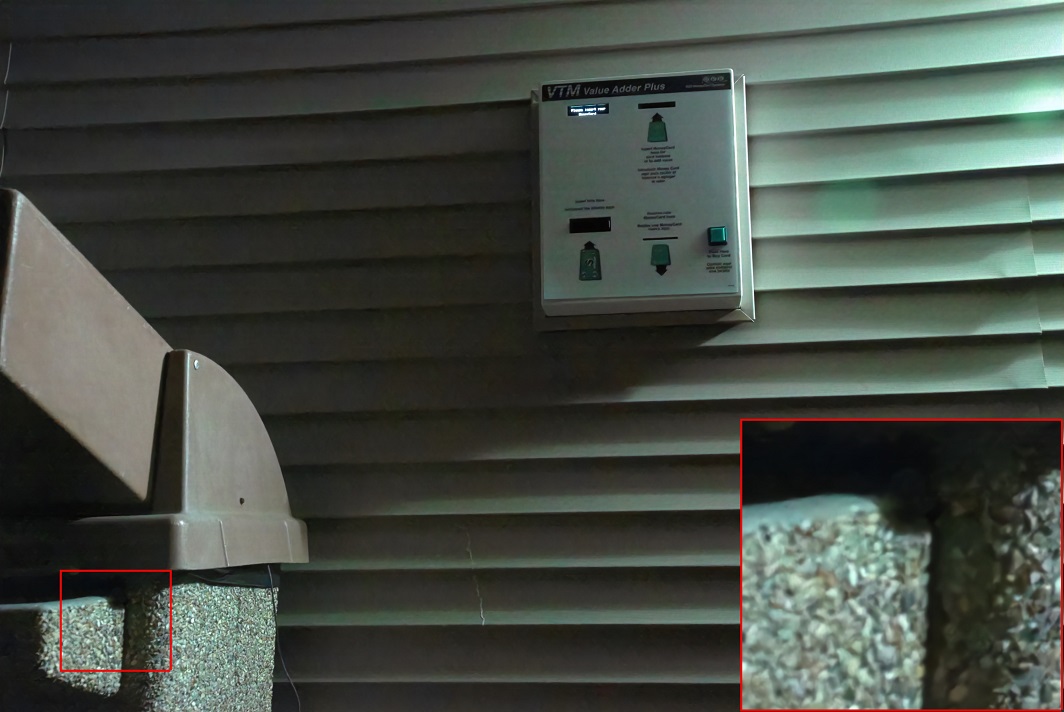}
   \includegraphics[width=1.0\textwidth]{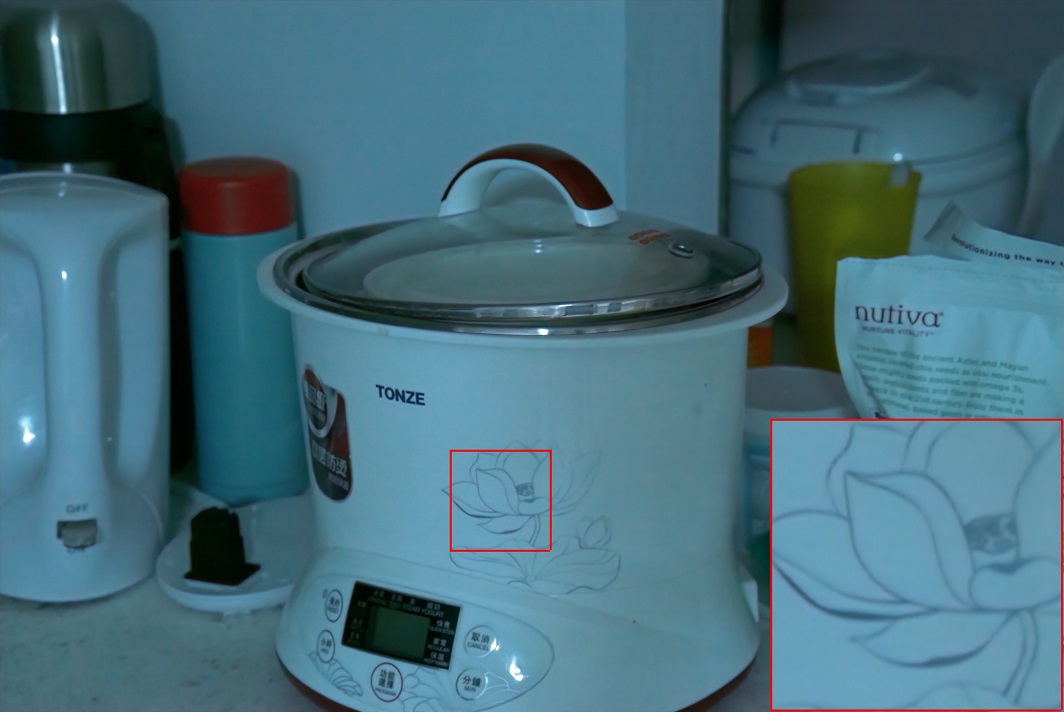}
   \centerline{(b) Residual }
  \end{minipage}
  \hfill
  \begin{minipage}[b]{0.24\linewidth}
   \centering
   \includegraphics[width=1.0\textwidth]{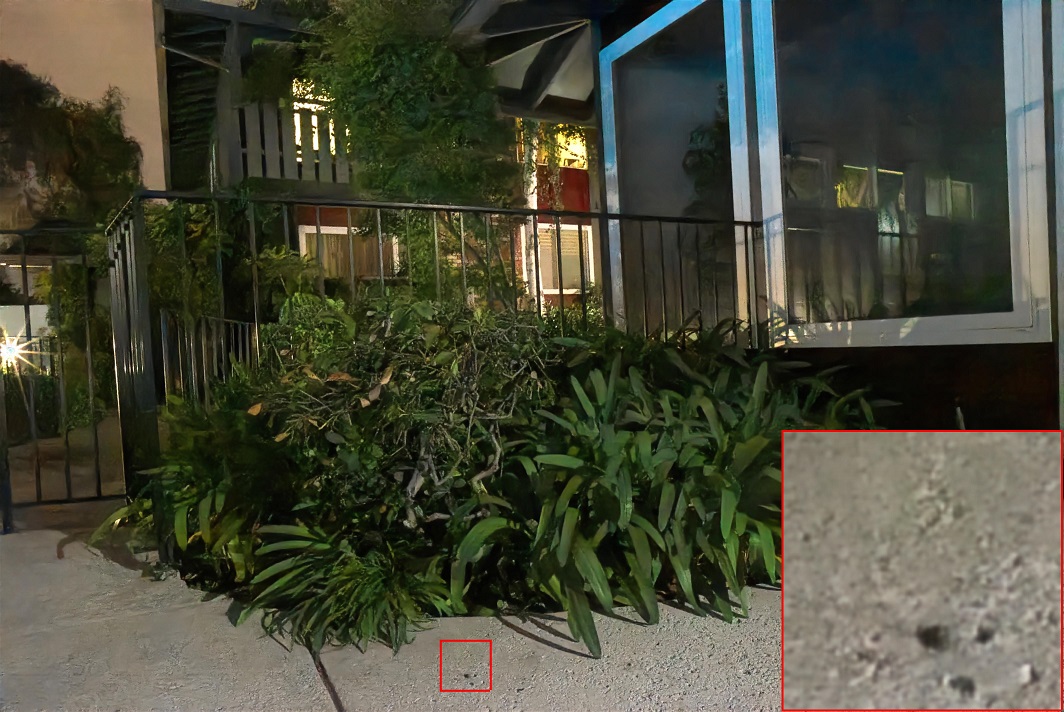}
   \includegraphics[width=1.0\textwidth]{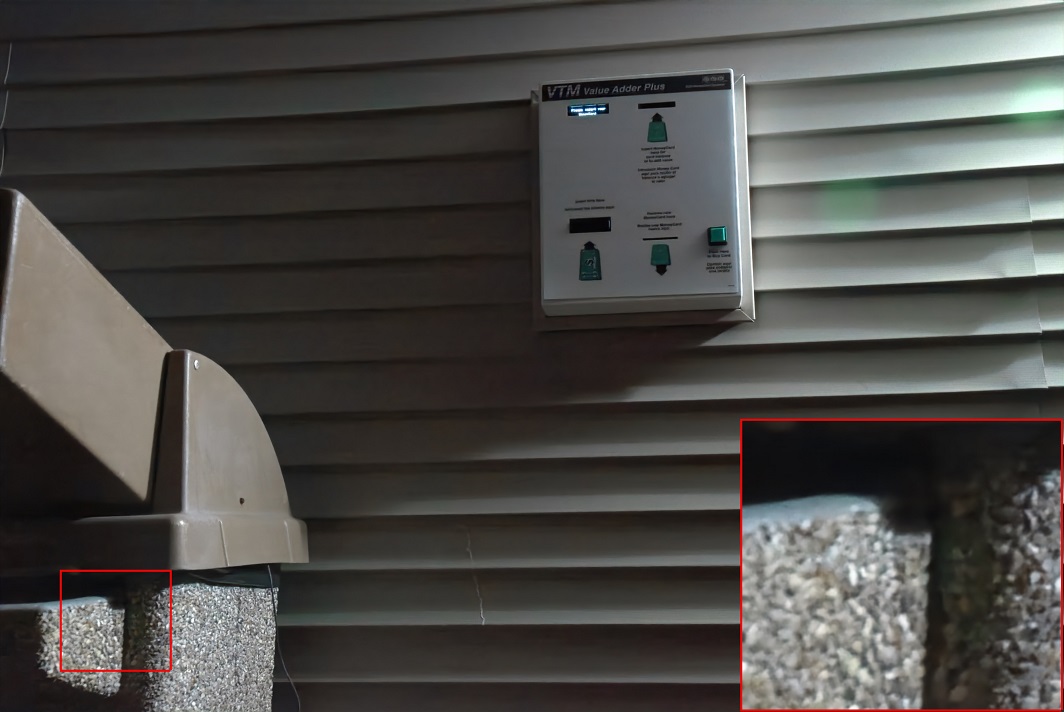}
   \includegraphics[width=1.0\textwidth]{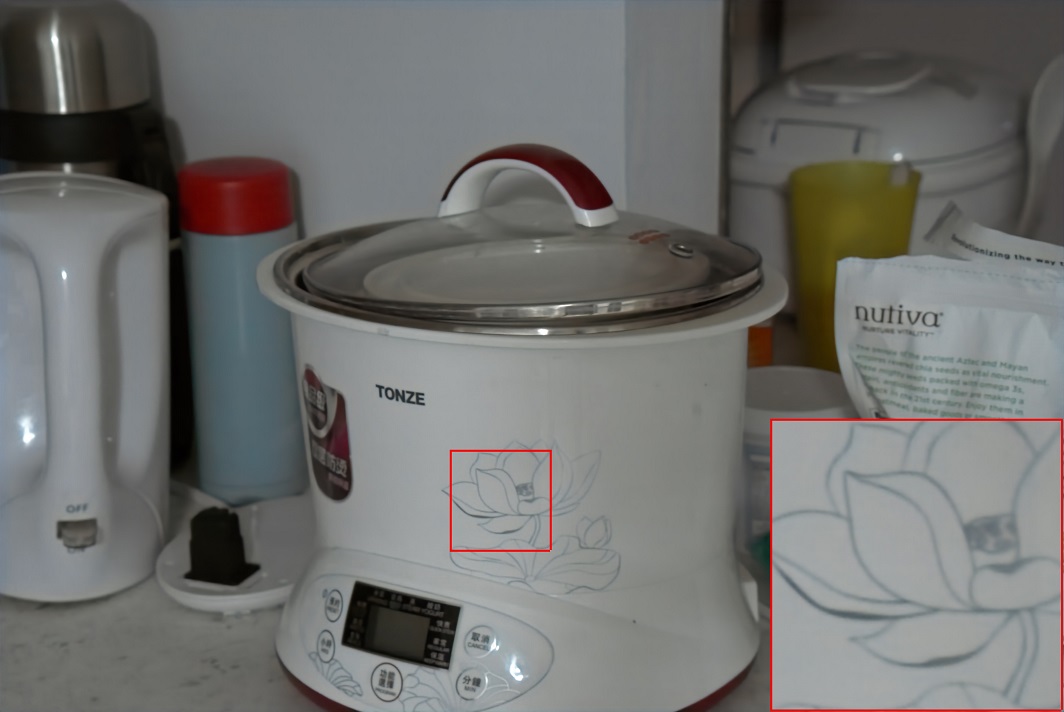}
   \centerline{(c) Ours}
  \end{minipage}
  \hfill
  \begin{minipage}[b]{0.24\linewidth}
   \centering
   \includegraphics[width=1.0\textwidth]{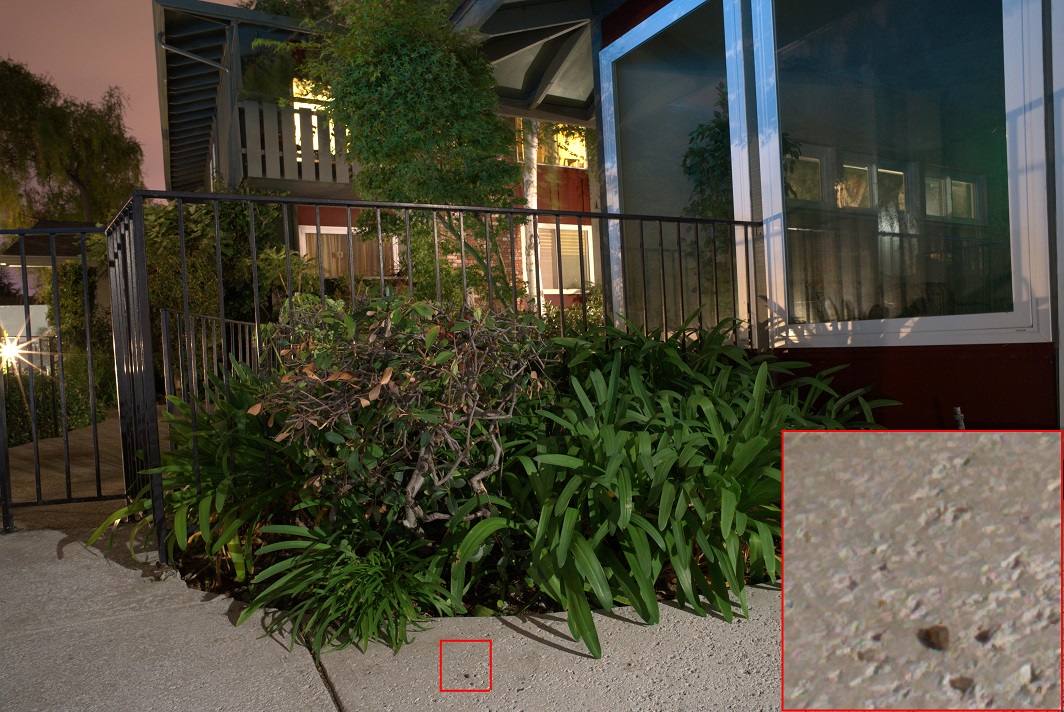}
   \includegraphics[width=1.0\textwidth]{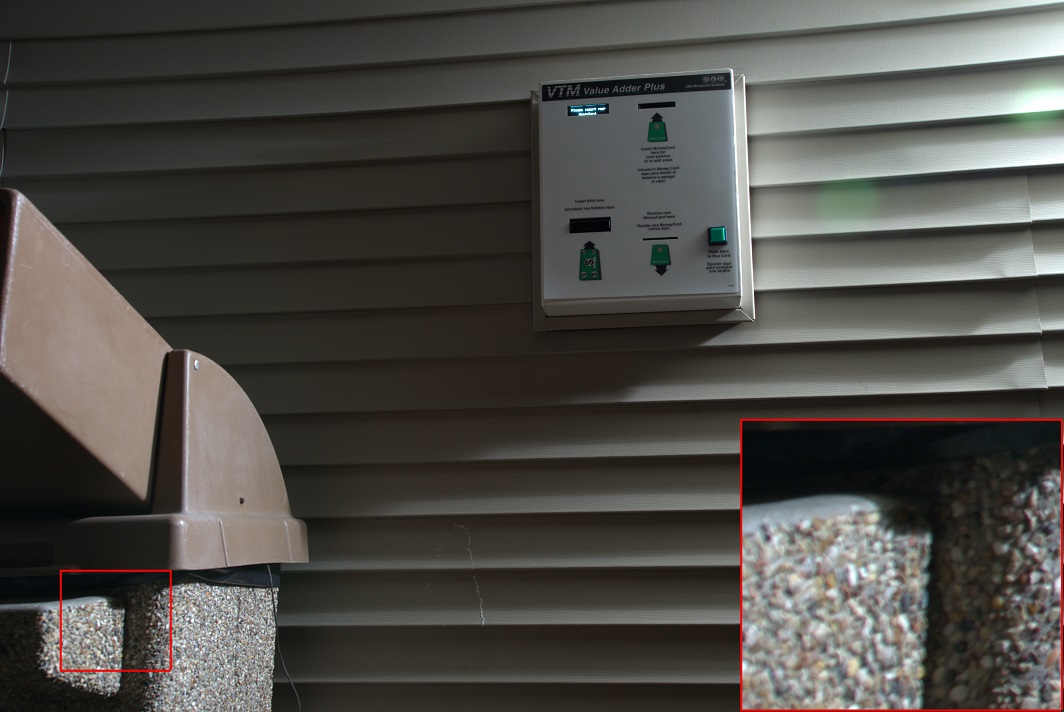}
   \includegraphics[width=1.0\textwidth]{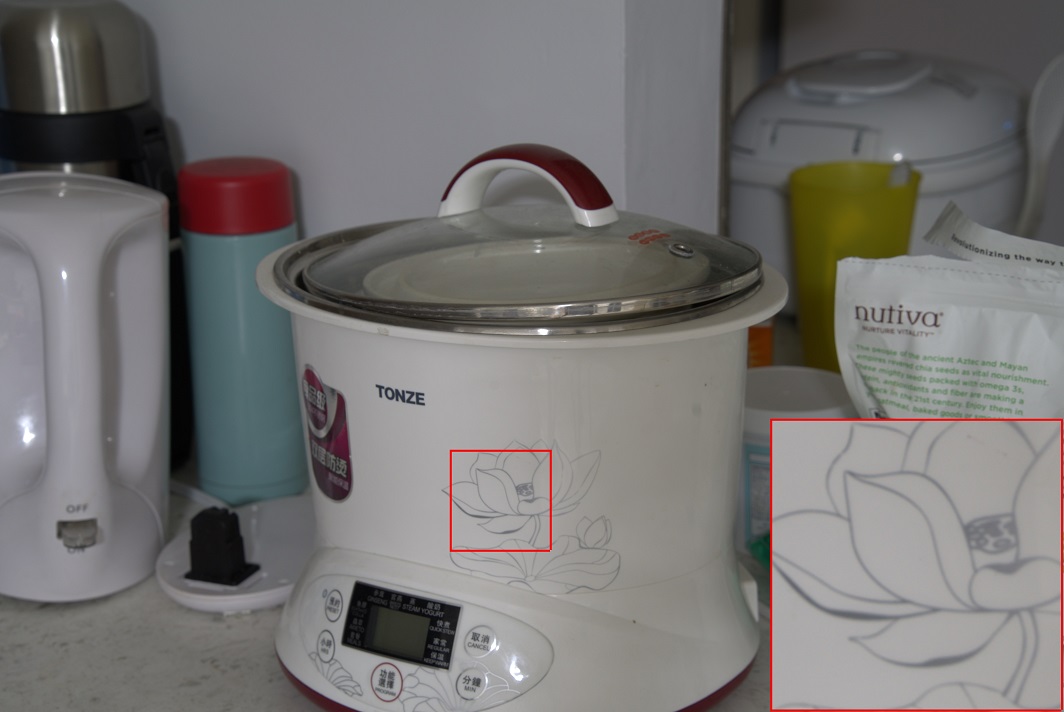}
   \centerline{(d) Ground truth}
  \end{minipage}
  \caption{Visual comparison with previous methods. Comparing to the top images, the results of our method have less noise. The middle and bottom images prove that our method can suppress the color spread. Please zoom in for a better view.
  }
  \label{fig:Comparison with others}
\end{figure*}

\section{EXPERIMENT}

\subsection{Dataset and Evaluation Metrics}

We adopt See-in-the-dark (SID) dataset \cite{chen2018learning} to evaluate the performance of our method. The SID dataset contains 5094 short-exposure images and 424 long-exposure images, which are raw sensor data captured by Sony $\alpha$7SII and Fujifilm X-T2 in extreme low-light environment. In this dataset, each scene has a sequence of images with different short-exposure time and a long-exposure image as a reference image. The short-exposure times were set between 0.033s and 0.1s. And the long-exposure times of corresponding reference images were set between 10s and 30s. In our experiments, we train and test our network with images captured by Sony camera, and employ PSNR and SSIM to evaluate the network performance for low-light image enhancement.

\subsection{Training and Testing}

We implemented our network with Pytorch and trained the network with 4000 epochs on the SID dataset. During the training, we used  Adam optimizer and set the initial learning rate to 0.0001. The learning rate decreased to 2e-5 after 2000 epochs and to 1e-5 after 3000 epochs. Before feeding to the network, we multiply the image patch by four amplification ratios, which provide multiple brightness images as input together. We set the amplification ratios as $w*\{0.5,0.8,1.0,1.2\}$, where $w$ represents the exposure difference between the input and reference images similar to \cite{chen2018learning}. In each iteration of training, we performed a random crop to get a $512\times 512$ patch from the raw image and flipped, rotated or transposed it randomly for data augmentation. The full images are taken as input in the testing for avoiding obvious boundary artifacts.
The entire network is conducted on a PC with NVIDIA Tesla V100 GPU with 32 GB of memory.

\subsection{Comparison with Other Methods}
We compared our method with the following methods, including BM3D \cite{dabov2007image}, SID \cite{chen2018learning}, Residual \cite{maharjan2019improving}. For the sake of fairness, we apply the code provided by the authors with recommended parameters setting. 

Comparing the results in Fig. \ref{fig:Comparison with others}, it can be observed that the quality of the enhanced images by our network is significantly higher than that of others. The method of SID \cite{chen2018learning} and Residual \cite{maharjan2019improving} generate incorrect color when it removes noise from low-light images. By visually comparison, we have noticed that our method has two improvements in enhancement with other methods. First, our method can recover more details and texture from low-light images with serious noise. As shown in red rectangle of Fig. \ref{fig:Comparison with others}, the images generated by our method look more smooth and satisfactory. Second, our method can restore correct and natural color and avoid color spreading, making the enhanced images more realistic and closer to ground truth. In quantitative comparison, we evaluate the performance of these methods using PSNR and SSIM. Table \ref{tab:comparison} illustrates the detailed comparison. Our method has achieved better performance in all subset with different amplification ratio while keeping a small number of parameters, which indicates the effectiveness of proposed method rather than the effect of network parameter.

\subsection{Ablation Study}
To validate the effectiveness of each component in our network, we performed several experiments and compared the results by adding blocks step by step. In these experiments, the hyper-parameters in the training process of each model were maintained, and all networks were trained 4000 epochs to reach the convergence state.

\begin{figure}[t]
  \centering
  \begin{minipage}[b]{0.49\linewidth}
    \includegraphics[width=1.0\textwidth]{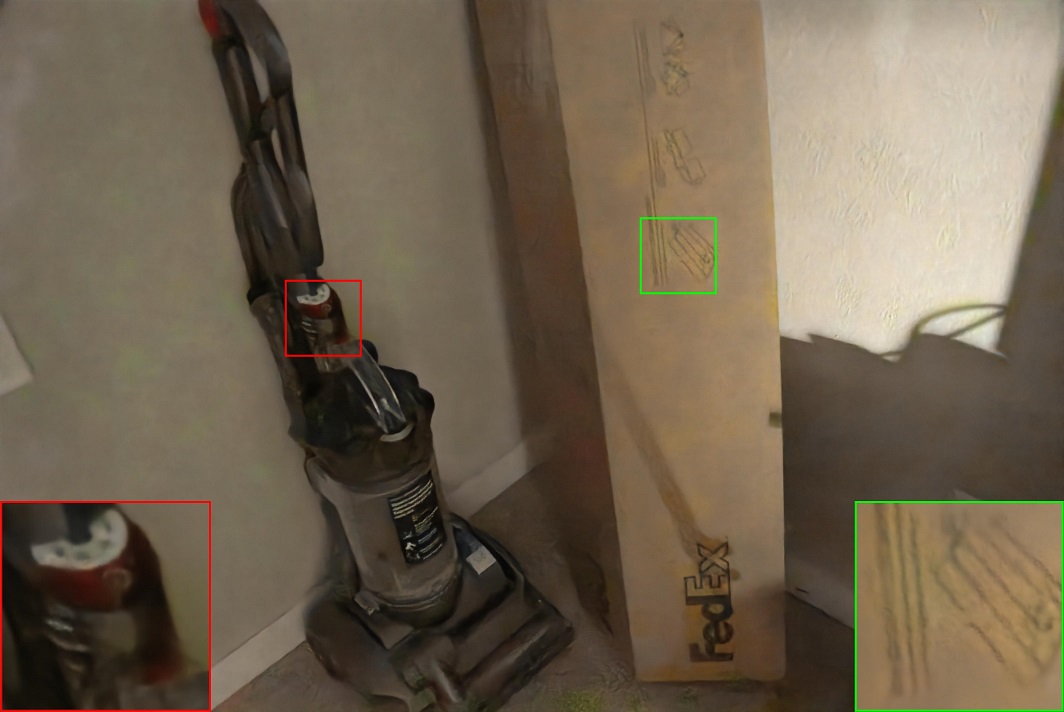}
    \centerline{(a) backbone}
  \end{minipage}
  \hfill
  \begin{minipage}[b]{0.49\linewidth}
    \includegraphics[width=1.0\textwidth]{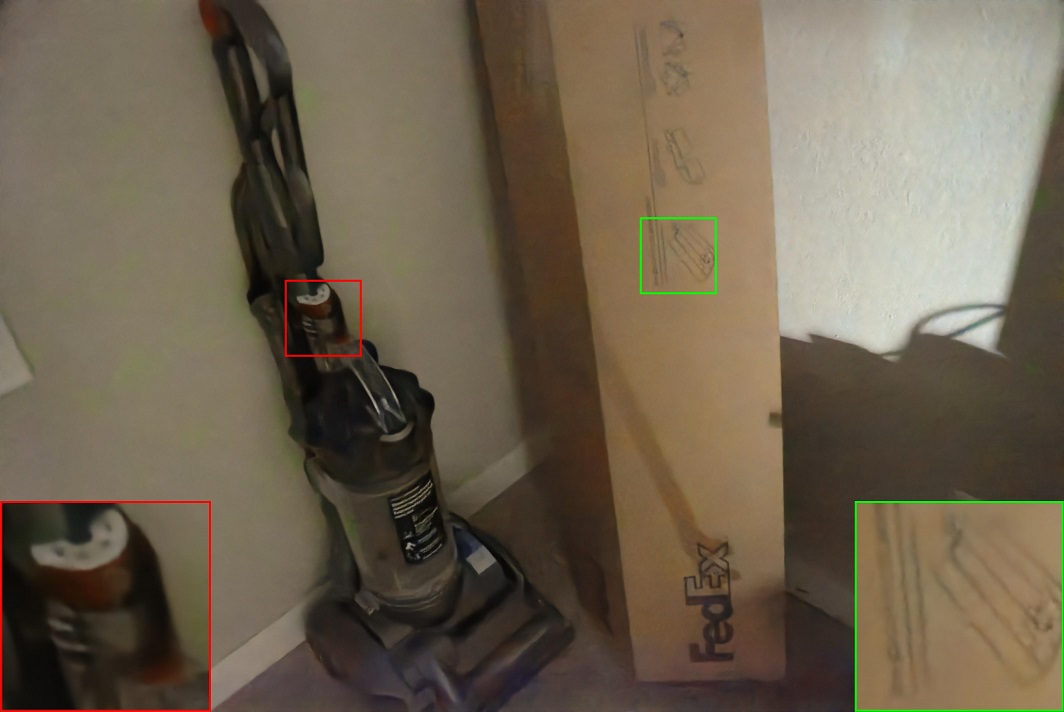}
    \centerline{(b) CAB }
  \end{minipage}
  \hfill
  \begin{minipage}[b]{0.49\linewidth}
    \includegraphics[width=1.0\textwidth]{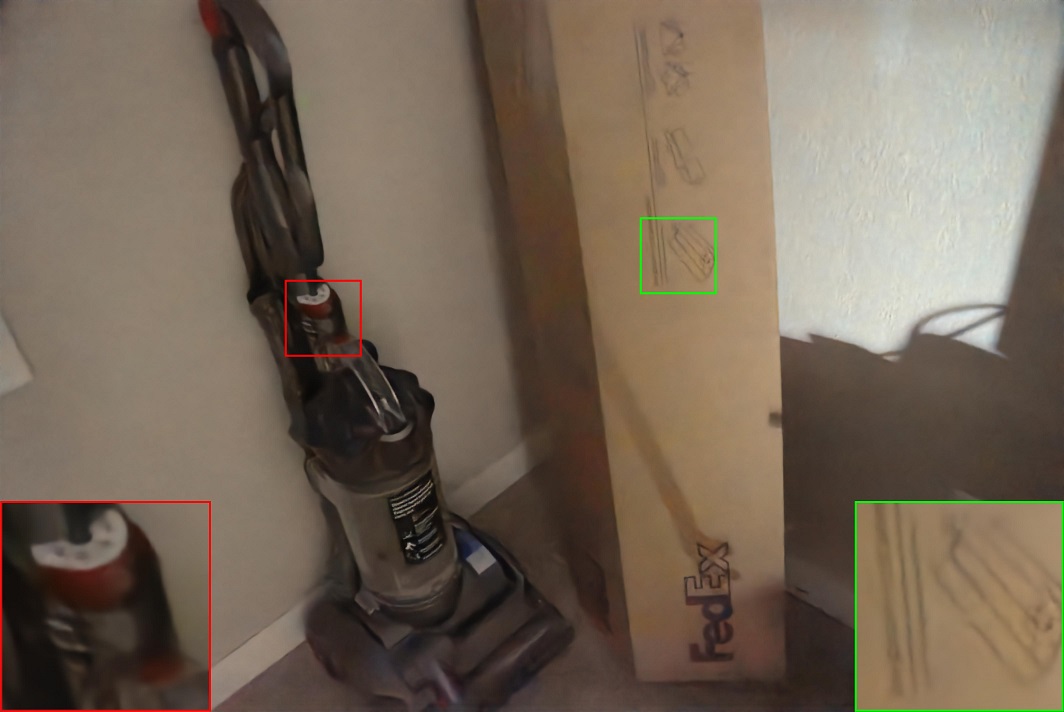}
    \centerline{(c) MAB}
  \end{minipage}
  \hfill
  \begin{minipage}[b]{0.49\linewidth}
    \includegraphics[width=1.0\textwidth]{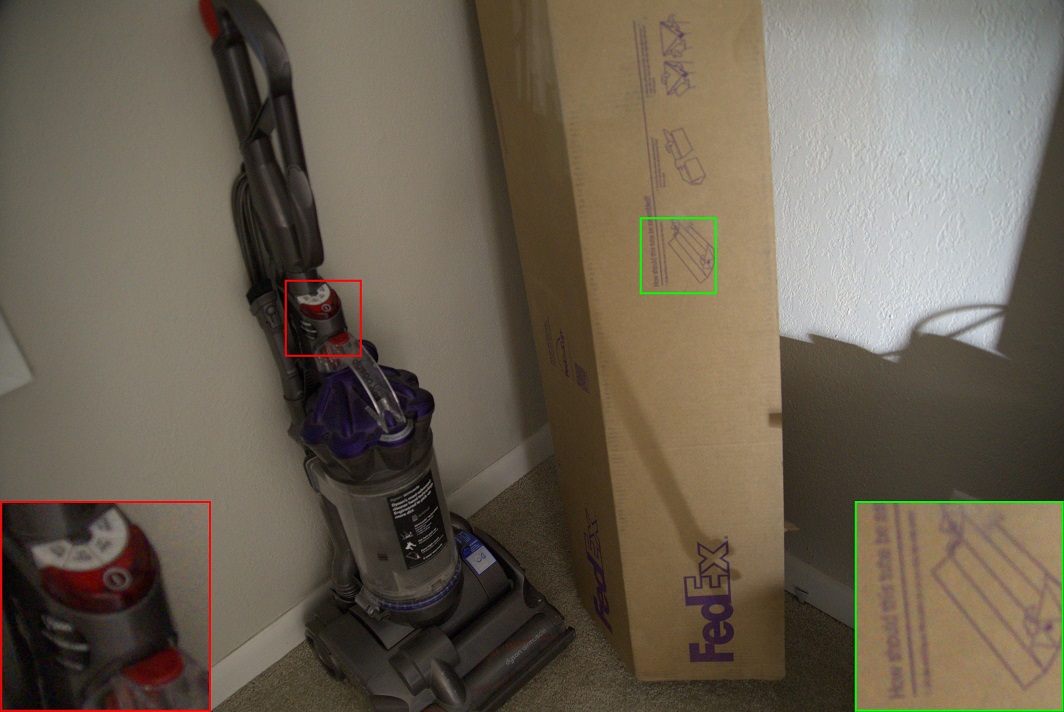}
    \centerline{(d) ground truth}
  \end{minipage}
  \caption{Visualisation of the impact with different modules. (a) is the result of backbone network. (b) and (c) are the outputs with CAB and MAB, respectively. (d) is the reference image.}
  \label{fig:Visualisation of modules}
\end{figure}

At first, we used a simple U-net structure similar to \cite{chen2018learning} as our backbone. Then CAB, MAB and ISL were added into the backbone one by one. We chose PSNR as an indicator to measure the impact of different modules on network performance. The results are shown in Table \ref{tab:study}. It can be seen from the comparison of the results in the table that CAB can significantly improve the PSNR of images. Figure \ref{fig:Visualisation of modules} illustrates the effect of adding different blocks. Through visual comparison, color artifacts and noise reduced greatly after adding these blocks, which means the blocks have a positive impact on improving image quality.

\begin{table}[t]
  \centering
  \caption{Quantitative comparison between our method and others on SID dataset. }
  ~\\
  \label{tab:comparison}
  \begin{tabular}{cc|cccc}
  \hline
    \multicolumn{2}{c|}{Method} &BM3D & SID  & Residual  & Ours \\
  \hline
    \multirow{4}*{PSNR} &x100 &18.76 &30.08 & 30.53& \textbf{31.53}\\
    &x250 &17.60 &28.42 &28.78 &\textbf{29.70}\\
    &x300 &17.42 &28.52 &28.38 &\textbf{28.66}\\
    &all & 17.88 &28.89 &29.16 &\textbf{29.86} \\
    \hline
    \multicolumn{2}{c|}{SSIM} & 0.498  &0.784 &0.783 &\textbf{0.787}\\
    \multicolumn{2}{c|}{Parameters} & - &7.76M &\textbf{2.5M} &2.6M\\
  \hline
  \end{tabular}
\end{table}

\begin{table}[t]
\begin{center}
\caption{Performance of network with different blocks.
} 
\label{tab:study}
\vspace{0.09cm}
\begin{tabular}{c|c|c|c|c}
 \hline
 backbone &CAB & MAB & ISL & PSNR 
 \\
 \hline
 $\surd$& &&& 28.57 \\
 $\surd$&$\surd$&& & 29.49 \\
 $\surd$&$\surd$&$\surd$&& 29.70 \\
 $\surd$&$\surd$&$\surd$&$\surd$& 29.86  \\
 \hline
\end{tabular}
\end{center}
\end{table}

\section{Conclusion and Future Work}

In this paper, we propose an attention-based network to enhance the raw images to obtain color images with high contrast and noiseless. Our method uses the mixed attention block with combining spatial and channel attention to extract features, making the network more efficiency. In addition, we use inverted shuffle layers instead of max pooling layers to retain more information. Experiments demonstrate that our method can generate enhanced images with less noise and color artifacts, achieving the best performance on the SID dataset.
In future work, we will explore a more effective attention module to decrease the computation cost and improve the network generalization ability.

%

\bibliographystyle{IEEEbib}
\bibliography{icme2020template}

\begin{thebibliography}{10}

\bibitem{guo2019toward}
Shi Guo, Zifei Yan, Kai Zhang, Wangmeng Zuo, and Lei Zhang,
\newblock ``Toward convolutional blind denoising of real photographs,''
\newblock in {\em Proceedings of the IEEE Conference on Computer Vision and
  Pattern Recognition}, 2019, pp. 1712--1722.

\bibitem{shen2017msr}
Liang Shen, Zihan Yue, Fan Feng, Quan Chen, Shihao Liu, and Jie Ma,
\newblock ``Msr-net: Low-light image enhancement using deep convolutional
  network,''
\newblock {\em arXiv preprint arXiv:1711.02488}, 2017.

\bibitem{chen2018image}
Jingwen Chen, Jiawei Chen, Hongyang Chao, and Ming Yang,
\newblock ``Image blind denoising with generative adversarial network based
  noise modeling,''
\newblock in {\em Proceedings of the IEEE Conference on Computer Vision and
  Pattern Recognition}, 2018, pp. 3155--3164.

\bibitem{chen2018learning}
Chen Chen, Qifeng Chen, Jia Xu, and Vladlen Koltun,
\newblock ``Learning to see in the dark,''
\newblock in {\em Proceedings of the IEEE Conference on Computer Vision and
  Pattern Recognition}, 2018, pp. 3291--3300.

\bibitem{maharjan2019improving}
Paras Maharjan, Li~Li, Zhu Li, Ning Xu, Chongyang Ma, and Yue Li,
\newblock ``Improving extreme low-light image denoising via residual
  learning,''
\newblock in {\em 2019 IEEE International Conference on Multimedia and Expo
  (ICME)}. IEEE, 2019, pp. 916--921.

\bibitem{guo2016lime}
Xiaojie Guo, Yu~Li, and Haibin Ling,
\newblock ``Lime: Low-light image enhancement via illumination map
  estimation,''
\newblock {\em IEEE Transactions on image processing}, vol. 26, no. 2, pp.
  982--993, 2016.

\bibitem{dabov2007image}
Kostadin Dabov, Alessandro Foi, Vladimir Katkovnik, and Karen Egiazarian,
\newblock ``Image denoising by sparse 3-d transform-domain collaborative
  filtering,''
\newblock {\em IEEE Transactions on image processing}, vol. 16, no. 8, pp.
  2080--2095, 2007.

\bibitem{gu2014weighted}
Shuhang Gu, Lei Zhang, Wangmeng Zuo, and Xiangchu Feng,
\newblock ``Weighted nuclear norm minimization with application to image
  denoising,''
\newblock in {\em Proceedings of the IEEE conference on computer vision and
  pattern recognition}, 2014, pp. 2862--2869.

\bibitem{zhang2017beyond}
Kai Zhang, Wangmeng Zuo, Yunjin Chen, Deyu Meng, and Lei Zhang,
\newblock ``Beyond a gaussian denoiser: Residual learning of deep cnn for image
  denoising,''
\newblock {\em IEEE Transactions on Image Processing}, vol. 26, no. 7, pp.
  3142--3155, 2017.

\bibitem{yan2020deep}
Qingsen Yan, Lei Zhang, Yu~Liu, Yu~Zhu, Jinqiu Sun, Qinfeng Shi, and Yanning
  Zhang,
\newblock ``Deep hdr imaging via a non-local network,''
\newblock {\em IEEE Transactions on Image Processing}, vol. 29, pp. 4308--4322,
  2020.

\bibitem{yan2019attention}
Qingsen Yan, Dong Gong, Qinfeng Shi, Anton van~den Hengel, Chunhua Shen, Ian
  Reid, and Yanning Zhang,
\newblock ``Attention-guided network for ghost-free high dynamic range
  imaging,''
\newblock {\em arXiv preprint arXiv:1904.10293}, 2019.

\bibitem{gong2018learning}
Dong Gong, Zhen Zhang, Qinfeng Shi, Anton van~den Hengel, Chunhua Shen, and
  Yanning Zhang,
\newblock ``Learning an optimizer for image deconvolution,''
\newblock {\em arXiv preprint arXiv:1804.03368}, 2018.

\bibitem{wei2018deep}
Chen Wei, Wenjing Wang, Wenhan Yang, and Jiaying Liu,
\newblock ``Deep retinex decomposition for low-light enhancement,''
\newblock {\em arXiv preprint arXiv:1808.04560}, 2018.

\bibitem{wang2019underexposed}
Ruixing Wang, Qing Zhang, Chi-Wing Fu, Xiaoyong Shen, Wei-Shi Zheng, and Jiaya
  Jia,
\newblock ``Underexposed photo enhancement using deep illumination
  estimation,''
\newblock in {\em Proceedings of the IEEE Conference on Computer Vision and
  Pattern Recognition}, 2019, pp. 6849--6857.

\bibitem{liang2019cameranet}
Zhetong Liang, Jianrui Cai, Zisheng Cao, and Lei Zhang,
\newblock ``Cameranet: A two-stage framework for effective camera isp
  learning,''
\newblock {\em arXiv preprint arXiv:1908.01481}, 2019.

\bibitem{hu2018squeeze}
Jie Hu, Li~Shen, and Gang Sun,
\newblock ``Squeeze-and-excitation networks,''
\newblock in {\em Proceedings of the IEEE conference on computer vision and
  pattern recognition}, 2018, pp. 7132--7141.

\bibitem{wang2018non}
Xiaolong Wang, Ross Girshick, Abhinav Gupta, and Kaiming He,
\newblock ``Non-local neural networks,''
\newblock in {\em Proceedings of the IEEE Conference on Computer Vision and
  Pattern Recognition}, 2018, pp. 7794--7803.

\bibitem{shi2016real}
Wenzhe Shi, Jose Caballero, Ferenc Husz{\'a}r, Johannes Totz, Andrew~P Aitken,
  Rob Bishop, Daniel Rueckert, and Zehan Wang,
\newblock ``Real-time single image and video super-resolution using an
  efficient sub-pixel convolutional neural network,''
\newblock in {\em Proceedings of the IEEE conference on computer vision and
  pattern recognition}, 2016, pp. 1874--1883.

\end{thebibliography}

\end{document}